\documentclass[final,3p,times,twocolumn,fixfloat]{elsarticle}

\usepackage[caption=false]{subfig}
\usepackage{graphicx}
\usepackage{epsfig,slashed,nicefrac}
\usepackage{color}
\usepackage{float} 
\usepackage{subfig} 
\usepackage{multirow}
\usepackage{amssymb} 
\usepackage{amsmath} 
\usepackage{times}
\usepackage{txfonts}
\usepackage{verbatim}

\journal{Physics Letters B} 

\usepackage{ifpdf}
 \ifpdf
 \usepackage[pdftex]{hyperref}
 \else
 \usepackage[hypertex]{hyperref}
 \fi

 \hypersetup{
   pdftitle={},%
   pdfauthor={},%
   pdfsubject={},%
   pdfkeywords={},%
   pdfstartview={},%
   bookmarksopen=true, breaklinks=true, debug=true, %
   colorlinks=true, linkcolor=red, citecolor=blue, urlcolor=blue
 }

\newcommand{\amc}{{\sc MadGraph5\textunderscore}a{\sc MC@NLO}}
\newcommand{\fr}{{\sc Feyn\-Rules}}

\newcommand{\ml}{{\sc MadLoop}}
\newcommand{\mfks}{{\sc MadFKS}}
\newcommand{\mw}{{\sc MadWidth}}
\newcommand{\ms}{{\sc MadSpin}}
\newcommand{\py}{{\sc Pythia}}
\newcommand{\nloct}{{\sc NloCT}}

\newcommand{\ninja}{{\sc Ninja}}
\newcommand{\fj}{{\sc FastJet}}

\chardef\MyArticleWithColor=\pdfcolorstackinit page direct{0 g}

\def\be{\begin{equation*}}
\def\ee{\end{equation*}}
\def\bsp#1\esp{\begin{split}#1\end{split}} 
\def\bpm{\begin{pmatrix}}
\def\epm{\end{pmatrix}}

\begin{document}
\begin{frontmatter}

\title{NLO predictions for the production of a spin-two particle at the LHC}

\author[SINP]{Goutam Das}
\author[DU]{C\'eline Degrande}
\author[SLAC]{Valentin Hirschi}
\author[UCL]{Fabio Maltoni}
\author[CERN]{Hua-Sheng Shao}

\address[SINP]{Theory Division, Saha Institute of Nuclear Physics, 1/AF Bidhan Nagar, Kolkata 700 064, India}
\address[DU]{
  Institute for Particle Physics Phenomenology,
  Department of Physics Durham University, Durham DH1 3LE,
  United Kingdom}
\address[SLAC]{SLAC, National Accelerator Laboratory,
  2575 Sand Hill Road, Menlo Park, CA 94025-7090, USA}
\address[UCL]{Centre for Cosmology, Particle Physics and Phenomenology (CP3),
 Universit\'e catholique de Louvain}
\address[CERN]{Theoretical Physics Department, CERN, CH-1211 Geneva 23, Switzerland}

\date{\today}

\begin{abstract}
\noindent We obtain  predictions accurate at the next-to-leading order in QCD for the production of a  generic spin-two particle in the most relevant channels at the LHC:  production in association with coloured particles (inclusive, one jet, two jets and $t\bar t$), with  vector bosons ($Z,W^\pm,\gamma$) and with the Higgs boson.  We present  total and differential cross sections as well as branching ratios as a function of the mass and the collision energy also considering the case of non-universal couplings to standard model particles. We find that the 
next-to-leading order corrections give rise to sizeable $K$ factors for many channels, in some cases exposing the unitarity-violating behaviour of non-universal couplings scenarios, and in general greatly reduce the theoretical uncertainties. Our predictions are publicly available in the \amc\ framework and can, therefore, be directly used in experimental simulations of spin-two particle production for arbitrary values of the mass and couplings.  
\end{abstract}


\end{frontmatter}

\section{Introduction}

After the discovery of the 125 GeV Higgs boson at the LHC~\cite{Aad:2012tfa,Chatrchyan:2012ufa},  the main task of Run II is to explore higher energy scales searching for physics beyond the Standard Model (SM). Evidence for new physics could be gathered via accurate measurements of the interactions among SM particles or from the detection of new particles. The existence of new particles at the TeV scale is widely motivated by both theoretical and experimental issues of the SM. While  no significant evidence for new resonances has been reported  at the LHC so far, searches are actively pursued by the experimental collaborations with approaches that are as model independent as possible~\cite{CMS:exotica,ATLAS:exotica}. 
For example, heavy colour-singlet states of arbitrary spins are searched for in several decay channels, including very clean ones (such as dilepton and diphoton) as well as more challenging ones, from diboson ($WW,ZZ,HZ, HH$) to  di-jet (with or without $b$-tags) and $t \bar t$ signatures.  Finally, associated production with SM particles are also often considered. 
 
Robust interpretations of the corresponding experimental bounds obtained on rates ($\sigma \cdot$ BR) need model assumptions on the one hand and  accurate and precise predictions for the cross sections and decay rates, on the other.  Most of the interpretations for spin-0 and spin-1 models are based on next-to-leading and next-to-next-to-leading order predictions, as these can be easily obtained by generalising SM calculations performed for the Higgs boson (in the SM or SUSY) and for the vector bosons. 

Interpretations for spin-two resonances, however, are typically performed via leading order computations, which due to their low accuracy and precision lead to a systematic loss in reach. A complementary limitation also exists for dedicated spin-two searches in the context of the many theoretical models predict the presence of massive spin-two resonances. The Kaluza-Klein excitations of the graviton and the composite bound state from strong dynamics are well-known examples of such scenarios. In this case, having accurate predictions can improve the experimental selections and significantly increase the sensitivity of the searches. In addition, having predictions at hand for other production mechanisms or decay modes can provide ideas for new signatures to be looked for, especially in the case of the detection of a signal. 

The aim of this Letter is to provide for the first time a complete implementation of the Lagrangian of a generic spin-two particle so that all the relevant production channels for the LHC can be accurately simulated at Next-to-Leading Order (NLO) in QCD.  In this context, accurate predictions and in particular event generators at least at NLO  in QCD and matched to Parton Showers (PS) are necessary to obtain  simulations that can directly be used by the experimental collaborations to allow information to be efficiently extracted from experimental data. While predictions for generic classes of bosonic resonances have become available in the last years, e.g.~\cite{Artoisenet:2013puc} and several results are known in the literature~\cite{Das:2014tva,Mathews:2004xp,Kumar:2008pk,Kumar:2009nn,Agarwal:2009xr,Gao:2009pn,Karg:2009xk,deAquino:2011ix,Frederix:2012dp,Frank:2012wh,Frederix:2013lga,Greiner:2013gca}, a completely general setup for the calculation at NLO in QCD of processes involving  a spin-two particle has still been lacking. Especially, NLO results with PS effects are new for almost all processes presented here, where only the inclusive spin-two particle production in the universal coupling case is in exception. Moreover, $2\rightarrow 3$ processes computed here are achieved at NLO accuracy for the first time in this Letter, while other processes like $Y_2+H/Z/W$ are also first available in the warped dimensional models by taking into account the QCD corrections. We stress that although the discovery itself could not need such an accurate Monte Carlo simulation, the characterisation of  a new state, from the determination of its quantum numbers to the form and strength of its couplings, will require the best predictions to be available to the experimental community. 

\medskip

\section{Theoretical framework}

We consider the effective field theory of a massive spin-two particle $Y_2$ interacting with the SM fields. The kinetic term of $Y_2$ can be described by the well-known Fierz-Pauli Lagrangian, with the positive-energy condition $\partial_\mu Y_2^{\mu\nu}=0$, and  the interactions with SM fields are  ($V$ is a gauge field, while $f$ are matter fields  )
\be\bsp
& {\cal L}^{Y_2}_{\rm V,f} = -\frac{\kappa_{V,f}}{\Lambda}T_{\mu\nu}^{V,f} Y_2^{\mu\nu}\,,
\esp \ee
where $T_{\mu\nu}^V$ ($T_{\mu\nu}^f$) are the energy-momentum tensors of $V$ ($f$), respectively, {\it  i.e.}, 
\be\bsp
 & T^V_{\mu\nu} =
      -g_{\mu\nu}\left[-\frac{1}{4}F^{\rho\sigma}F_{\rho\sigma}+\delta_{m_V,0}\left(\left(\partial^{\rho}\partial^{\sigma}V_{\sigma}\right)V_{\rho}+\frac{1}{2}\left(\partial^{\rho}V_{\rho}\right)^2\right)\right]\\
  &\quad
-F_{\mu}^{\rho}F_{\nu\rho}+\delta_{m_V,0}\left[\left(\partial_{\mu}\partial^{\rho}V_{\rho}\right)V_{\nu}
+\left(\partial_{\nu}\partial^{\rho}V_{\rho}\right)V_{\mu}\right]\,,\\
  &T^f_{\mu\nu} =
    -g_{\mu\nu}\left[\bar{\psi}_f\left(i\gamma^{\rho}D_{\rho}-m_f\right)\psi_f-\frac{1}{2}\partial^{\rho}\left(\bar{\psi}_fi\gamma_{\rho}\psi_f\right)\right]\\
  &\quad
   +\left[\frac{1}{2}\bar{\psi}_fi\gamma_{\mu}D_{\nu}\psi_f-\frac{1}{4}\partial_{\mu}\left(\bar{\psi}_fi\gamma_{\nu}\psi_f\right)+\left(\mu\leftrightarrow \nu\right)\right]\,,
\esp \ee
where  the indices of other possible quantum numbers (such as colour) are understood and $F_{\mu\nu}$ is the field strength of $V$. In the SM, the gauge fields $V$ are ${\rm SU(2)}_L\times {\rm U(1)}_Y$ ElectroWeak (EW) gauge bosons ($W,B$) or the ${\rm SU(3)}_C$ gluon $g$, while the matter fields $f$ are quarks, leptons and left-handed neutrinos. The gauge-fixed term proportional to the Kronecker delta function $\delta_{m_V,0}$ in $T^V_{\mu\nu}$ indicates that it is needed only when $V$ is massless $m_V=0$ ({\it i.e.}, $V=g,\gamma$). The $Y_2$ can also interact with the SM Higgs doublet $\Phi$  via
\be\bsp
& {\cal L}^{Y_2}_{\rm \Phi} = -\frac{\kappa_H}{\Lambda}T_{\mu\nu}^{\Phi}Y_2^{\mu\nu}\,,
\esp\ee
where the energy-momentum tensor $T_{\mu\nu}^{\Phi}$ 
is
\be\bsp
T_{\mu\nu}^{\Phi}=D_{\mu}\Phi^{\dagger}D_{\nu}\Phi+D_{\nu}\Phi^{\dagger}D_{\mu}\Phi-g_{\mu\nu}(D^{\rho}\Phi^{\dagger}D_{\rho}\Phi-V(\Phi))\,.
\esp\ee
After spontaneous symmetry breaking, one gets the mass eigenstates of EW bosons ($Z,W^{\pm},\gamma$) and SM Higgs boson $H$. In addition, when working in the Feynman gauge and at 1-loop level, the extra interaction of $Y_2$ and Fadeev-Popov (FP) ghost fields is necessary~(e.g. Refs.~\cite{Mathews:2004pi,Coriano:2011zk}),
\be\bsp
& {\cal L}^{Y_2}_{\rm FP} = -\frac{\kappa_V}{\Lambda}T_{\mu\nu}^{\rm FP}Y_2^{\mu\nu}\,,
\esp\ee
where
\be\bsp
& T^{\rm FP}_{\mu\nu} = -g_{\mu\nu}\left[\left(\partial^{\rho}\bar{\omega}^a\right)\left(\partial_{\rho}\omega^a\right)-g_sf^{abc}\left(\partial^{\rho}\bar{\omega}^a\right)\omega^bV^c_{\rho}\right]\\
&\quad
+\left[\left(\partial_{\mu}\bar{\omega}^a\right)\left(\partial_{\nu}\omega^a\right)-g_sf^{abc}\left(\partial_{\mu}\bar{\omega}^a\right)\omega^bV^c_{\nu}+\left(\mu\leftrightarrow \nu\right)\right],
\esp\ee
 $\omega$ being the FP ghost  of the gluon field $V=g$ and $g_s$ the strong coupling constant.

Our implementation builds upon the \fr\ package~\cite{Christensen:2008py,Alloul:2013bka} and the \nloct\ program~\cite{Degrande:2014vpa} which are used to generate the UFO model~\cite{Degrande:2011ua} as well as the counterterms for  the renormalisation and the rational term $R_2$. Some extended functionalities have been implemented in \nloct\  to handle the effective Lagrangian of a spin-two particle. A point worth of stressing concerns the renormalisation.  With universal couplings, {\it e.g}, $\kappa_g=\kappa_q$  no extra renormalisation procedure is needed beyond the usual ones of the SM as the spin-two current is conserved. On the contrary, for non-universal  couplings, the spin-two current is not conserved and specific renormalisation constants need to be introduced to cancel left-over ultraviolet divergences~\cite{Artoisenet:2013puc}. These extra couplings are renormalised as 
\be\bsp
& \delta \kappa_g = \frac{\alpha_s}{3\pi}T_F\sum_q{\left(\kappa_g-\kappa_q\right)}\left(\frac{1}{\epsilon}-\gamma_E+\log{4\pi}+\log{\frac{\mu_R^2}{m_{Y_2}^2}}\right)\,,\\
& \delta \kappa_q = \frac{2\alpha_s}{3\pi}C_F\left(\kappa_q-\kappa_g\right)\left(\frac{1}{\epsilon}-\gamma_E+\log{4\pi}+\log{\frac{\mu_R^2}{m_{Y_2}^2}}\right)\,,
\esp\ee
by  \nloct, where $C_F=\frac{4}{3},T_F=\frac{1}{2}$. Our implementation is general and allows for models with non-universal couplings case to be studied at NLO accuracy. The finite part of these counterterms identifies the renormalisation scheme where the couplings $\kappa_{g,q}$ are defined as $\kappa_{g,q}(m_{Y_2})$ and it is chosen so that these couplings do not run at this order in perturbation theory.

The corresponding  spin-two UFO model~\cite{FR-NLOmodels} 
is directly employable in the \amc\ framework~\cite{Alwall:2014hca} to perform phenomenological studies at NLO QCD accuracy including matching to PS. One-loop contributions are calculated numerically by the \ml\ module~\cite{Hirschi:2011pa} with the tensor integrand-level reduction method~\cite{Ossola:2006us,Hirschi:2016mdz} that was implemented in \ninja~\cite{Mastrolia:2012bu,Peraro:2014cba}. The real emission contributions are calculated with the Frixione-Kuntz-Signer (FKS) subtraction method~\cite{Frixione:1995ms,Frixione:1997np} implemented in \mfks~\cite{Frederix:2009yq}. Finally, the MC@NLO formalism~\cite{Frixione:2002ik} is employed to perform the matching between fixed-order NLO calculations and PS, hence making event generation possible.


\begin{figure*}[t]
\centering
\vspace{-3.5cm}
\includegraphics[width=1.0\textwidth]{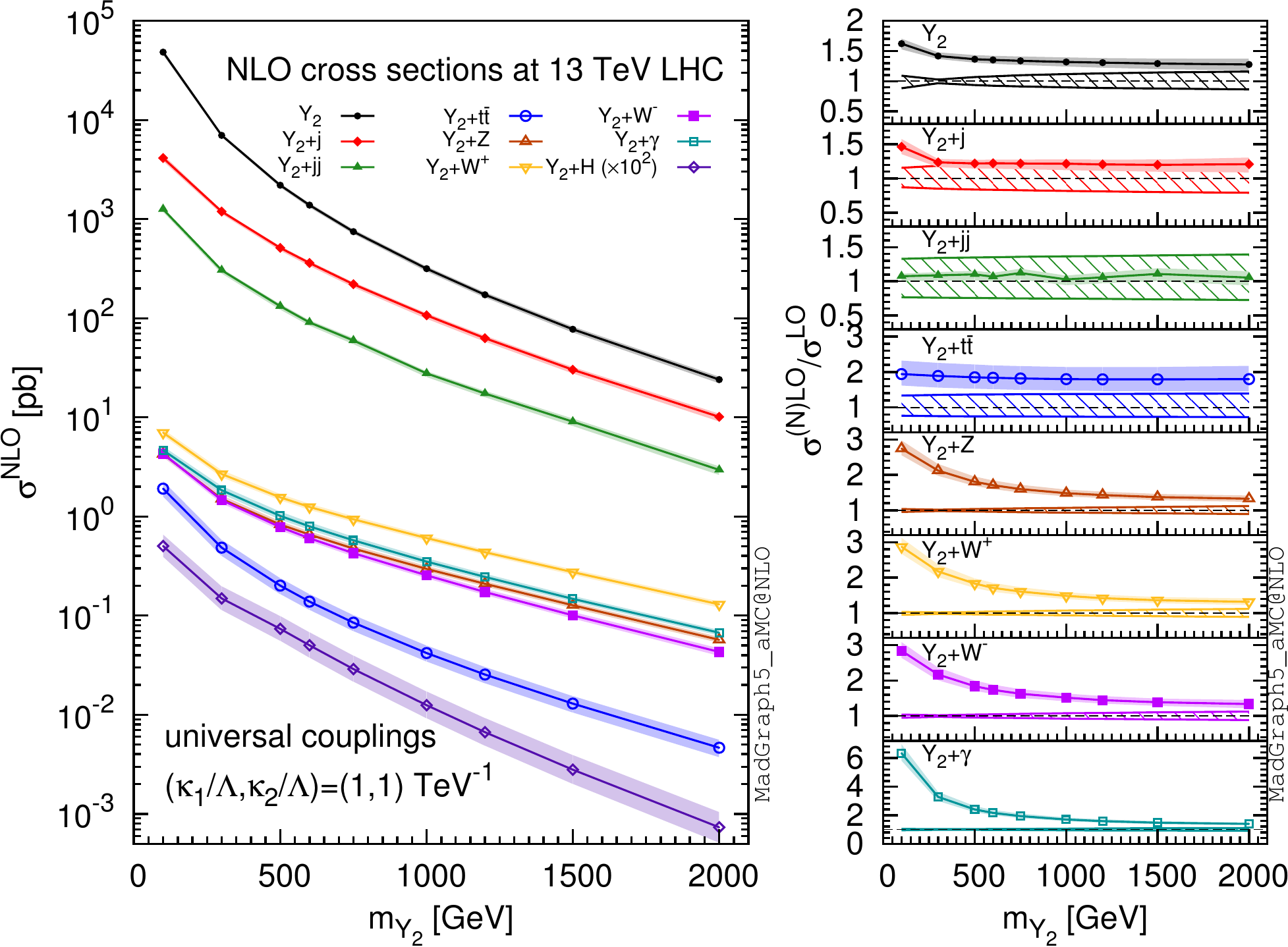}
  \caption{\small \label{fig:tot_xs}Summary plot of NLO cross sections and corresponding K factors for the various spin-two particle production processes listed in Table 1, where the scale, PDF and $\alpha_s$ uncertainties have been taken into account.}
\end{figure*}


\section{Production at LHC}

\begin{figure*}[t]
\centering
\vspace{-5cm}
\includegraphics[width=1.1\textwidth]{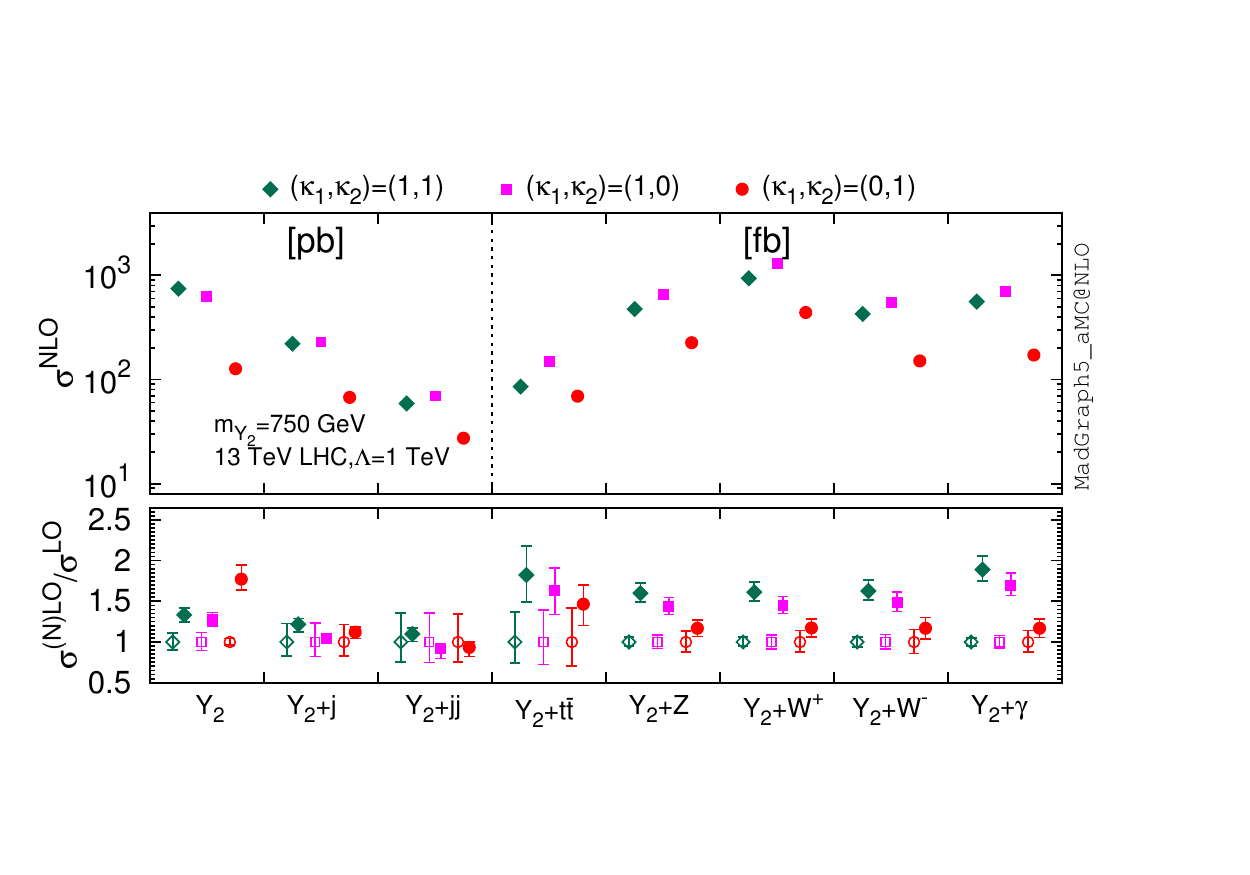}
  \caption{ \small \label{fig:tot_xs2}Summary plot of NLO cross sections and corresponding K factors for the various spin-two particle production processes in both universal and nonuniversal coupling cases at $m_{Y_2}=750$ GeV, where the scale, PDF and $\alpha_s$ uncertainties have been taken into account in $\frac{\sigma^{\rm (N)LO}}{\sigma^{\rm LO}}$ (lower panel).}
\end{figure*}

We now present predictions for the production of a  spin-two particle $Y_2$ as a function of mass as well centre-of-mass energy at a hadron collider, for a wide range of production channels. We will then focus on the LHC with a center-of-mass energy of \mbox{$\sqrt{s}=$13~TeV}. The (N)LO total cross sections of various $Y_2$ production processes in the universal coupling case (i.e. $\frac{\kappa_i}{\Lambda}=1~{\rm TeV}^{-1}$) are given in Table~\ref{tot_xs} for 500 GeV, 750 GeV and 1 TeV resonance masses and summarised in Figure~\ref{fig:tot_xs}. We also consider the minimal ``basis''  of predictions, the universal couplings ($(\frac{\kappa_1}{\Lambda},\frac{\kappa_2}{\Lambda})=(1,1)~{\rm TeV}^{-1}$). The non-universal couplings cases  ($(\frac{\kappa_1}{\Lambda},\frac{\kappa_2}{\Lambda})=(1,0), (0,1)~{\rm TeV}^{-1}$), where the definition of $\kappa_1$ and $\kappa_2$ are given in Table~\ref{couplings}, are discussed later for the intermediate reference mass point of 750 GeV, see Figure~\ref{fig:tot_xs2}.

We have employed NLO PDF4LHC15~\cite{Butterworth:2015oua,Dulat:2015mca,Harland-Lang:2014zoa,Ball:2014uwa,Gao:2013bia,Carrazza:2015aoa,Watt:2012tq} set with 30+2 members to estimate the PDF and $\alpha_s$ uncertainties. Missing higher-order QCD corrections are estimated by independently varying the renormalisation scale $\mu_R$ and factorization scale $\mu_F$ between $1/2 \mu_0$ to $2 \mu_0$, $\mu_0$ being the half of sum of the transverse masses of  the final states. In Table~\ref{tot_xs}, the quoted uncertainties come from scale variation, PDF and $\alpha_s$, respectively. Relevant SM parameters are the top mass $m_t=173.3$ GeV, the $Z$-boson mass $m_Z=91.1876$ GeV, the $W^{\pm}$ mass $m_W=79.82436$ GeV, the electromagnetic coupling constant $\alpha^{-1}(m_Z)=127.9$, and zero widths for all particles. For simplicity, we adopt the 5-flavour scheme and the CKM mixing matrix set to unity.

Cross sections for  i) $pp\rightarrow Y_2+j$, ii) $pp\rightarrow Y_2+jj$ and iii) $pp\rightarrow Y_2+\gamma$ require a jet (or photon) definition and kinematical cuts.  The jets are defined by the anti-$k_T$ algorithm~\cite{Cacciari:2008gp} as implemented in \fj~\cite{Cacciari:2011ma} with $R=0.4$. We also require  cuts on the transverse momentum $p_T(j)$ and the pseudorapidity $\eta(j)$ of jets. The photon is required to be isolated using Frixione's criterion~\cite{Frixione:1998jh}, where the isolation parameters used in Eq.~(3.4) of Ref.~\cite{Frixione:1998jh} have been set to $\epsilon_{\gamma}=1,n=1,\delta_0=0.4$. Cuts are chosen on a process-dependent basis: i) $p_T(j)>100$ GeV, ii) $p_T(j)>50$ GeV and $|\eta(j)|<4.5$ and $M(j_1,j_2)>400$ GeV, iii) $p_T(\gamma)>50$ GeV and $|\eta(\gamma)|<2.5$.

Several sources of theoretical uncertainties have been considered.  As expected, the PDF and the parametrical $\alpha_s$ uncertainties strongly depend on the process. $\sigma(pp\rightarrow Y_2+t\bar{t})$ suffers from the largest  PDF uncertainty, $7\%-8\%$, which is comparable in size to the scale uncertainty and due to the relatively poor knowledge of the gluon PDFs at large values of the Bjorken $x$.  $\sigma(pp\rightarrow Y_2+t\bar{t})$ and $\sigma(pp\rightarrow Y_2+jj)$, starting at order $\alpha_s^2$, are also sensitive to the $\alpha_s$ parametric uncertainty, while for   all other processes it is negligible. The scale uncertainties in the QCD processes ({\it i.e.}, $pp\rightarrow Y_2+j,Y_2+jj,Y_2+t\bar{t}$) are significantly reduced after including NLO corrections as expected. We also find that the estimate of the uncertainties at LO is not reliable for the EW processes ({\it i.e.}, $pp\rightarrow Y_2+Z,Y_2+W^{\pm},Y_2+\gamma$). 
For the sake of completeness, we have also computed the production of $Y_2$ in association with the Higgs boson $H$. Neglecting the bottom Yukawa coupling,  no tree-level diagrams appear and the leading contribution to $pp\rightarrow Y_2+H$ comes from the top-quark Loop-Induced (LI) diagrams. Exploiting the techniques of Ref.~\cite{Hirschi:2015iia},  these contributions can be automatically calculated at LO in \amc. As expected, the resulting cross section $\sigma(pp\rightarrow Y_2+H)$ is quite small compared those of the other processes.

The results in the universal coupling case are presented in Table~\ref{tot_xs} and Figure~\ref{fig:tot_xs} as a function of the resonance mass. They show that the K factors for the EW processes are larger for lower $m_{Y_2}$ masses. This can be accounted for by the new contributions coming from gluon-quark initial states that appear only beyond LO and whose importance increase at low Bjorken $x$. In the right panel of Figure~\ref{fig:tot_xs} LO uncertainty bands are included and represented by the hatched regions. It is interesting to note that the (N)LO uncertainty bands for these processes do not overlap, indicating the limitations of LO computations and the necessity of including QCD corrections for a reliable Monte Carlo simulation. 

\begin{table*}[t]
\hspace{0cm}
 \begin{tabular}{c|c|ccc}
 \hline
 $m_{Y_2}$ [GeV]   & Process & $\sigma^{\rm NLO}$~[pb] & $\sigma^{\rm LO}$~[pb] & K factor \\
\hline\hline
 & $pp\rightarrow Y_2$ & $(2.19\times 10^{3})^{+4.9\%}_{-5.5\%}\pm2.0\%$ &  $(1.60\times 10^{3})^{+6.2\%}_{-5.9\%}\pm2.7\%$	&  $1.37$ \\
 & $pp\rightarrow Y_2+j$ & $(5.13\times 10^{2})^{+2.9\%}_{-5.2\%}\pm3.1\%$ & $(4.21\times 10^2)^{+20.4\%}_{-15.9\%}\pm3.2\%$ & $1.22$\\
 & $pp\rightarrow Y_2+jj$ & $(1.33\times 10^{2})^{+2.4\%}_{-6.4\%}\pm4.3\%$ & $(1.21\times 10^2)^{+34.7\%}_{-24.0\%}\pm4.3\%$  & $1.10$ \\
 & $pp\rightarrow Y_2+t\bar{t}$ & $(2.01\times 10^{-1})^{+18.1\%}_{-16.2\%}\pm6.6\%$ & $(1.08\times 10^{-1})^{+35.8\%}_{-24.5\%}\pm6.2\%$ & $1.86$\\
$ 500 $  & $pp\rightarrow Y_2+Z$ & $(8.31\times 10^{-1})^{+8.0\%}_{-6.4\%}\pm1.8\%$ & $(4.60\times 10^{-1})^{+2.8\%}_{-2.9\%}\pm2.2\%$ & $1.81$\\
& $pp\rightarrow Y_2+W^+$ & $(1.56\times 10^{0})^{+8.0\%}_{-6.4\%}\pm1.9\%$ & $(8.52\times 10^{-1})^{+2.7\%}_{-2.8\%}\pm2.4\%$ & $1.77$\\
& $pp\rightarrow Y_2+W^-$ & $(7.81\times 10^{-1})^{+8.3\%}_{-6.6\%}\pm2.6\%$ & $(4.24\times 10^{-1})^{+2.8\%}_{-2.8\%}\pm2.9\%$ & $1.84$ \\
& $pp\rightarrow Y_2+\gamma$ & $(1.02\times 10^{0})^{+10.0\%}_{-8.0\%}\pm1.7\%$ & $(4.24\times 10^{-1})^{+1.9\%}_{-2.0\%}\pm2.2\%$ & $2.41$\\
& $pp\rightarrow Y_2+H$ (LI) & - & $(7.37\times 10^{-4})^{+34.0\%}_{-23.8\%}\pm5.2\%$ &  \\\hline
 & $pp\rightarrow Y_2$ & $(7.49\times 10^{2})^{+4.0\%}_{-4.0\%}\pm3.4\%$ &  $(5.59\times 10^{2})^{+9.2\%}_{-8.1\%}\pm3.4\%$	&  $1.34$ \\
 & $pp\rightarrow Y_2+j$ & $(2.20\times 10^{2})^{+2.8\%}_{-5.6\%}\pm3.9\%$ & $(1.81\times 10^2)^{+22.1\%}_{-16.9\%}\pm4.0\%$ & $1.22$\\
 & $pp\rightarrow Y_2+jj$ & $(5.97\times 10^{1})^{+2.9\%}_{-6.9\%}\pm4.9\%$ & $(5.33\times 10^1)^{+35.4\%}_{-24.4\%}\pm4.9\%$  & $1.12$ \\
 & $pp\rightarrow Y_2+t\bar{t}$ & $(8.50\times 10^{-2})^{+17.8\%}_{-16.2\%}\pm7.4\%$ & $(4.68\times 10^{-2})^{+36.4\%}_{-24.8\%}\pm6.9\%$ & $1.82$\\
$ 750 $  & $pp\rightarrow Y_2+Z$ & $(4.76\times 10^{-1})^{+7.1\%}_{-5.8\%}\pm2.4\%$ & $(2.98\times 10^{-1})^{+4.8\%}_{-4.5\%}\pm2.8\%$ & $1.60$\\
& $pp\rightarrow Y_2+W^+$ & $(9.38\times 10^{-1})^{+7.2\%}_{-5.8\%}\pm2.3\%$ & $(5.82\times 10^{-1})^{+4.7\%}_{-4.4\%}\pm2.8\%$ & $1.61$\\
& $pp\rightarrow Y_2+W^-$ & $(4.26\times 10^{-1})^{+7.4\%}_{-6.0\%}\pm3.3\%$ & $(2.62\times 10^{-1})^{+4.8\%}_{-4.5\%}\pm3.6\%$ & $1.63$ \\
& $pp\rightarrow Y_2+\gamma$ & $(5.74\times 10^{-1})^{+9.0\%}_{-7.2\%}\pm2.0\%$ & $(2.97\times 10^{-1})^{+4.0\%}_{-3.8\%}\pm2.3\%$ & $1.89$\\
& $pp\rightarrow Y_2+H$ (LI) & - & $(2.89\times 10^{-4})^{+35.7\%}_{-24.6\%}\pm6.3\%$ &  \\\hline
 & $pp\rightarrow Y_2$ & $(3.15\times 10^{2})^{+3.9\%}_{-4.2\%}\pm4.1\%$ &  $(2.39\times 10^{2})^{+11.1\%}_{-9.5\%}\pm4.0\%$	&  $1.32$ \\
 & $pp\rightarrow Y_2+j$ & $(1.07\times 10^{2})^{+2.7\%}_{-5.8\%}\pm4.7\%$ & $(8.81\times 10^1)^{+23.3\%}_{-17.7\%}\pm4.7\%$ & $1.22$\\
 & $pp\rightarrow Y_2+jj$ & $(2.78\times 10^{1})^{+2.0\%}_{-6.3\%}\pm5.1\%$ & $(2.70\times 10^1)^{+36.1\%}_{-24.7\%}\pm5.5\%$  & $1.03$ \\
 & $pp\rightarrow Y_2+t\bar{t}$ & $(4.20\times 10^{-2})^{+17.5\%}_{-16.1\%}\pm8.0\%$ & $(2.33\times 10^{-2})^{+36.8\%}_{-25.0\%}\pm7.5\%$ & $1.80$\\
$ 1000 $  & $pp\rightarrow Y_2+Z$ & $(2.96\times 10^{-1})^{+6.4\%}_{-5.3\%}\pm2.9\%$ & $(2.00\times 10^{-1})^{+6.3\%}_{-5.7\%}\pm3.4\%$ & $1.48$\\
& $pp\rightarrow Y_2+W^+$ & $(6.05\times 10^{-1})^{+6.4\%}_{-5.3\%}\pm2.8\%$ & $(4.07\times 10^{-1})^{+6.2\%}_{-5.6\%}\pm3.3\%$ & $1.49$\\
& $pp\rightarrow Y_2+W^-$ & $(2.55\times 10^{-1})^{+6.8\%}_{-5.6\%}\pm4.0\%$ & $(1.68\times 10^{-1})^{+6.3\%}_{-5.7\%}\pm4.4\%$ & $1.51$ \\
& $pp\rightarrow Y_2+\gamma$ & $(3.51\times 10^{-1})^{+7.9\%}_{-6.5\%}\pm2.2\%$ & $(2.07\times 10^{-1})^{+5.5\%}_{-5.1\%}\pm2.5\%$ & $1.70$\\
& $pp\rightarrow Y_2+H$ (LI) & - & $(1.25\times 10^{-4})^{+36.9\%}_{-25.3\%}\pm7.4\%$ &  \\\hline
\end{tabular}
\caption{\label{tot_xs}Total cross sections for various spin-two particle production processes with the effective field theory scale $\Lambda=1$ TeV. Results are presented together with the renormalisation/factorization scale, PDF$+\alpha_s$ uncertainties.}
\end{table*}

\begin{table}[t]
 \begin{center}
 \begin{tabular}{c|c}
    Process & Couplings set \\
\hline
 $pp\rightarrow Y_2, Y_2+j, Y_2+jj$ & $\kappa_1=\kappa_g,\kappa_2=\kappa_{q,t}$\\
 $pp\rightarrow Y_2+t\bar{t}$ & $\kappa_1=\kappa_{g,q},\kappa_2=\kappa_t$\\
 $pp\rightarrow Y_2+Z$ & $\kappa_1=\kappa_{g,q,t},\kappa_2=\kappa_{B,W,H}$\\
 $pp\rightarrow Y_2+W^{\pm}$ & $\kappa_1=\kappa_{g,q,t},\kappa_2=\kappa_{B,W,H}$\\
 $pp\rightarrow Y_2+\gamma$ & $\kappa_1=\kappa_{g,q,t},\kappa_2=\kappa_{B,W,H}$\\
 $pp\rightarrow Y_2+H$ & $\kappa_1=\kappa_{g,q,t},\kappa_2=\kappa_{B,W,H}$
\end{tabular}
\end{center}
\caption{\label{couplings}Definition of the couplings $\kappa_{1,2}$ for different processes.}
\end{table}

\begin{figure*}[t]
\centering
\hspace{-1cm}
  \subfloat[Transverse momentum distribution]{\includegraphics[width=1\columnwidth]{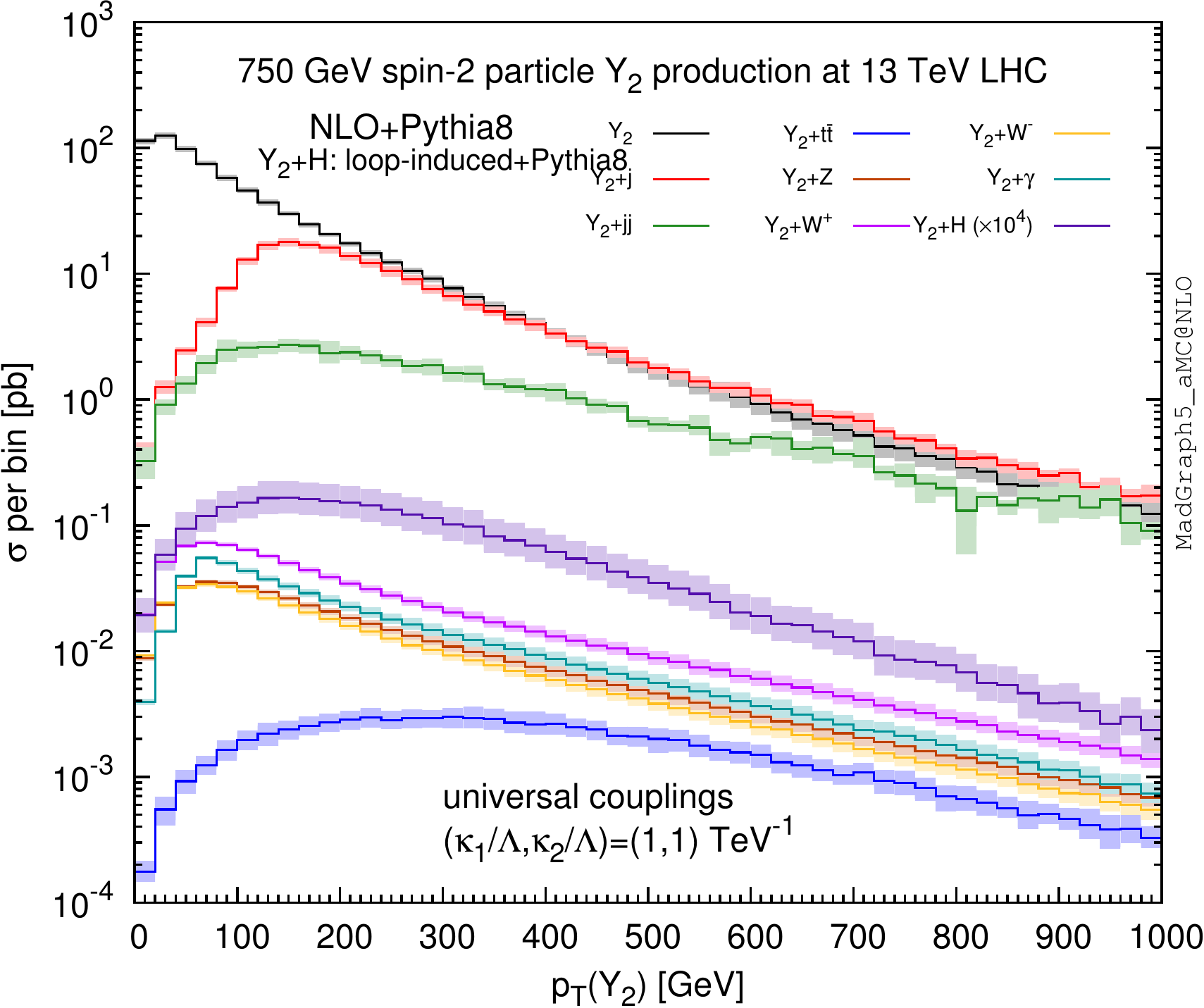}\label{fig:diff-a}}
\hspace{0.1 cm}
  \subfloat[Pseudorapidity distribution]{\includegraphics[width=1\columnwidth]{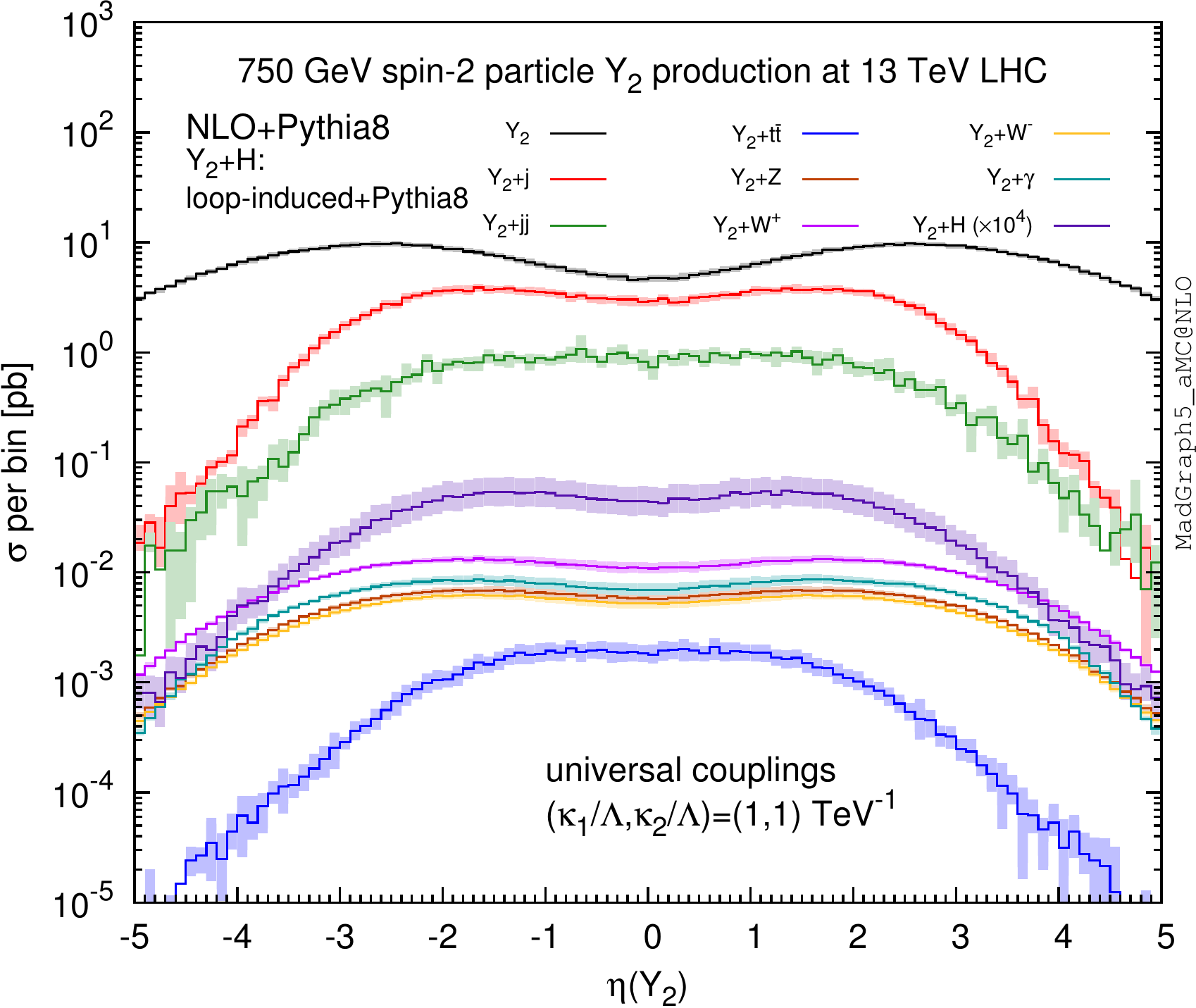}\label{fig:diff-b}}
  \caption{\small \label{fig:diff}Differential distributions  for various $Y_2$ production processes by matching NLO calculations with PS program \py\ 8.2 with the universal couplings assumption $(\kappa_1/\Lambda,\kappa_2/\Lambda)=(1,1)~{\rm TeV}^{-1}$: (a) transverse momentum spectrum of $Y_2$ (b) pseudorapidity distribution of $Y_2$. $Y_2+H$ is loop-induced and calculated at leading order. The error bands represent scale and PDF+$\alpha_s$ uncertainties.}
\end{figure*}

\begin{figure*}
\hspace{-0.2cm}
  \subfloat[]{\includegraphics[width=1.0\columnwidth]{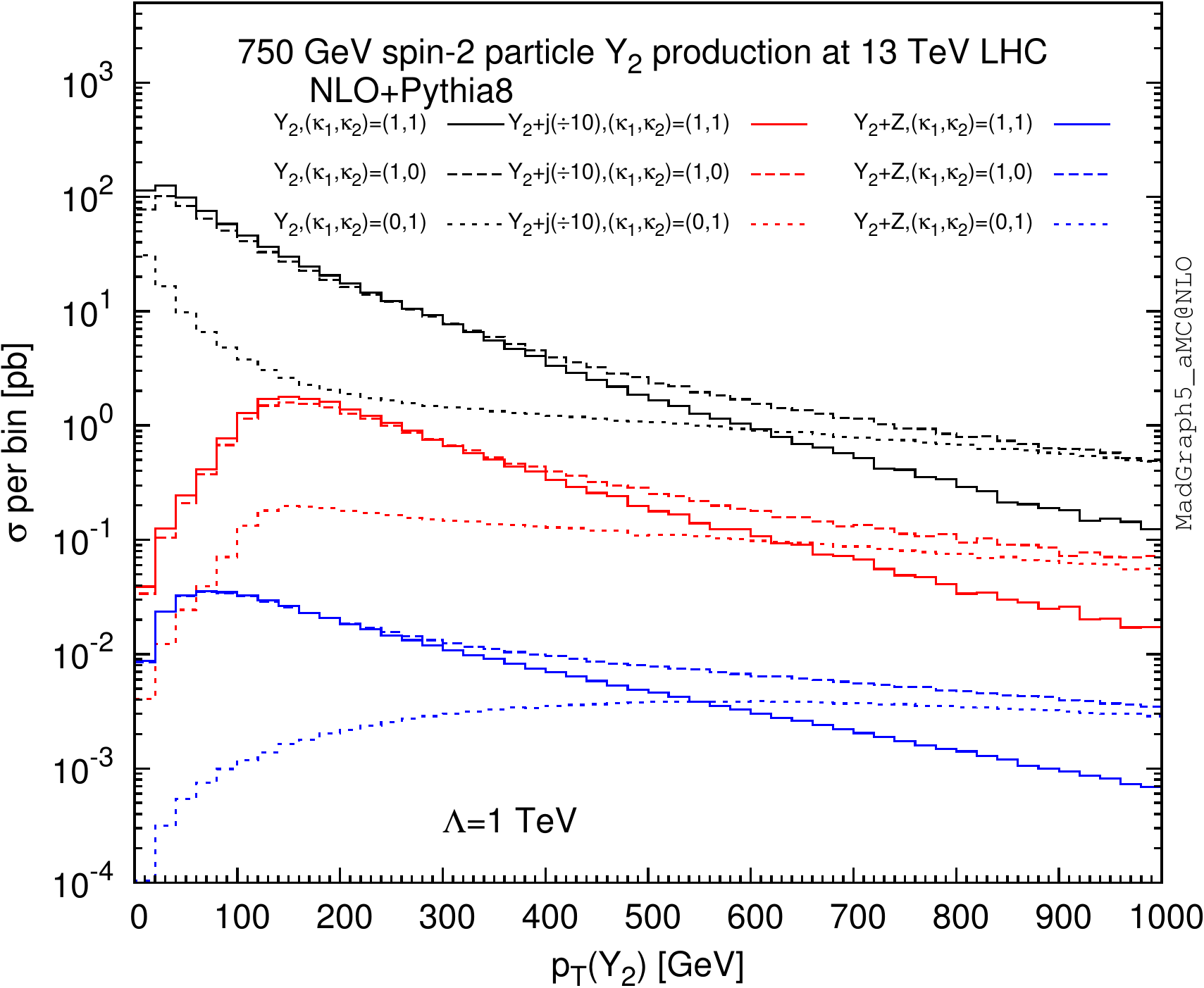}\label{fig:diff2-a}}
\hspace{0.1 cm}
  \subfloat[]{\includegraphics[width=1.0\columnwidth]{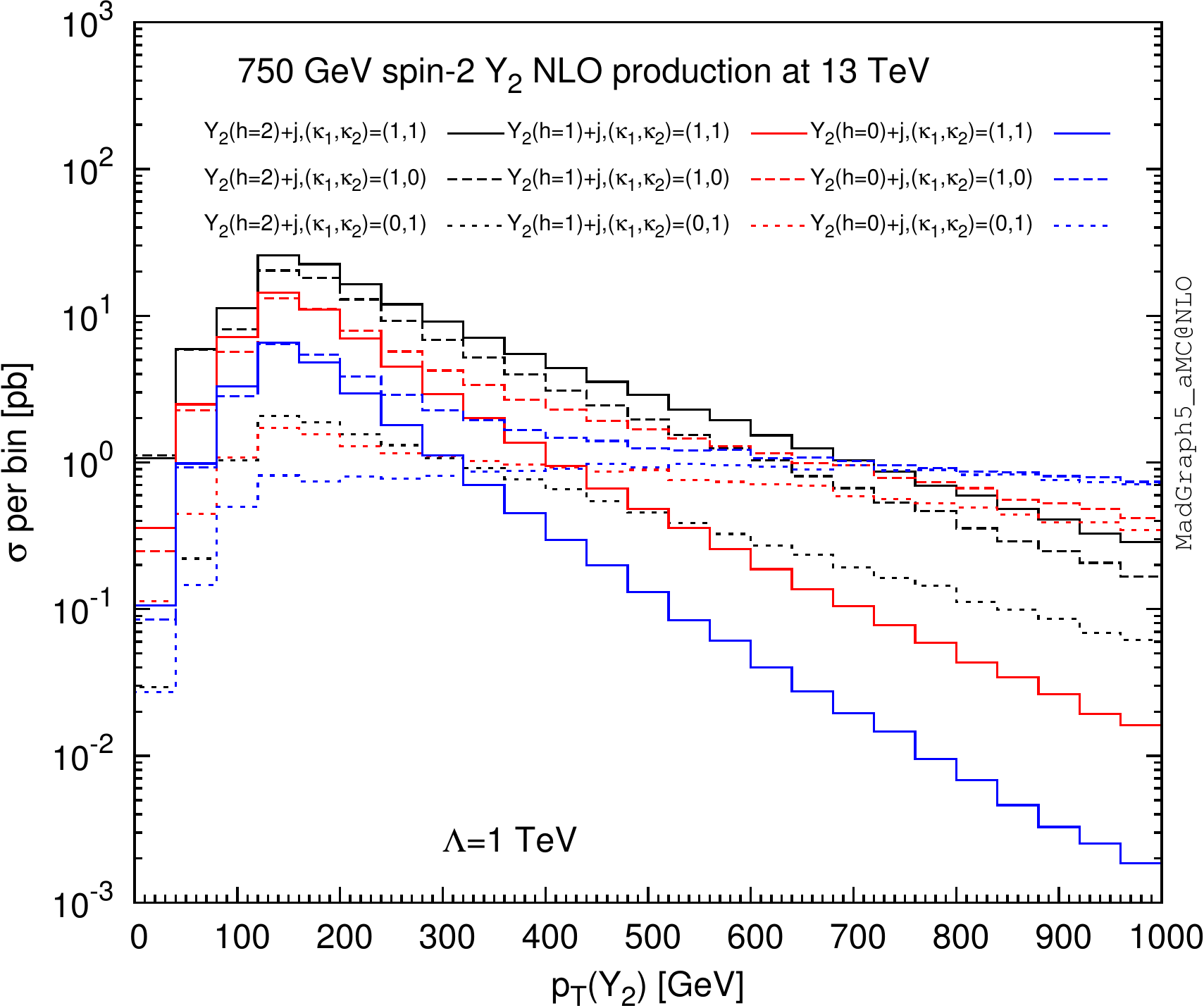}\label{fig:diff2-b}}
  \caption{\small \label{fig:diff2}Transverse momentum distributions of $Y_2$ with non-universal couplings (a) in $pp\rightarrow Y_2,Y_2+j,Y_2+Z$ at NLO accuracy matched to PS (b) in $pp\rightarrow Y_2+j$ with the breakdown of different $Y_2$ helicity contributions at NLO.}
\end{figure*}

The results shown in Figure~\ref{fig:tot_xs2} provide a useful ``basis" to evaluate cross sections for other choices of the couplings 
which can be written as  $\sigma(\kappa_1,\kappa_2) = \kappa_1^2 \sigma(1,0)+ \kappa_2^2 \sigma(0,1) +  \kappa_1 \kappa_2 (\sigma(1,1)- \sigma(1,0)-\sigma(0,1)) $. Note also that since the (N)LO cross sections for single $Y_2$ production processes are proportional to $\Lambda^{-2}$, one can fix  $\Lambda=1$ TeV and obtain results for other values of $\Lambda$ by a simple rescaling. 

The same codes that are used for the calculation of the total cross sections can also be employed as event generators at NLO accuracy by interfacing them to a PS program. Decays of the spin-two resonance can be included keeping spin correlations either by directly generating the corresponding process or by using the \ms\ package~\cite{Artoisenet:2012st}. Needless to say, out of a simulated sample, one can obtain any differential distribution of interest.  For the sake of illustration,  distributions after matching  NLO calculations  to \py\ 8.2~\cite{Sjostrand:2014zea} are presented in Figure~\ref{fig:diff} for the universal coupling case, {\it i.e.}, $(\kappa_1,\kappa_2)=(1,1)$, with $m_{Y_2}=750$ GeV as an example.  Figures \ref{fig:diff-a} and \ref{fig:diff-b} show the transverse momentum $p_T(Y_2)$ and pseudorapidity $\eta(Y_2)$ distributions of $Y_2$ for the nine production processes, respectively.   Note that for  inclusive production the accuracy of $\frac{d\sigma}{dp_T}(pp\rightarrow Y_2)$ is only at the LO level when $p_T(Y_2)\neq 0$. We stress that, even though we do not present the results here, inclusive samples with formal NLO accuracy for different jet multiplicities can be obtained by suitably merging NLO samples with the corresponding parton multiplicities. In \amc\ this  can be done automatically employing the FxFx method~\cite{Frederix:2012ps}. Nevertheless, one can already see that the curve $\frac{d\sigma}{dp_T}(pp\rightarrow Y_2+j)$ overlaps  with $\frac{d\sigma}{dp_T}(pp\rightarrow Y_2)$ when $p_T(Y_2)>400$ GeV. In this range $\frac{d\sigma}{dp_T}(pp\rightarrow Y_2+j)$ provides the NLO results for this observable and indeed one notices that the theoretical uncertainty is reduced. The differential $K$ factors in $p_T(Y_2)$ are rather constant for the three QCD processes, while they tend to increase with $p_T(Y_2)$ for the four EW processes. The increase in the latter case is due to  the opening of new partonic channels, quark-gluon initial states at NLO, while at LO only quark-antiquark initial states contribute.

We now turn to studying distributions for non-universal couplings cases. In Ref.~\cite{Artoisenet:2013puc} it was pointed out that when $\kappa_g \ne \kappa_q$ the spin-two current is not conserved and  $2\to 2$ squared amplitudes, such as 
$qg \to Y_2 q$, feature a dramatic growth with the parton level center of mass energy $\hat{s}$, scaling as  $(\kappa_g - \kappa_q)^2 \hat{s}^3/m_{Y_2}^4/\Lambda^2 $. We have reproduced the corresponding unitarity-violation curves for $pp\rightarrow Y_2$ with $m_{Y_2}=750$ GeV in Figure~\ref{fig:diff2-a}. In addition, we show the $p_T(Y_2)$ distributions in the non-universal coupling cases for $pp\rightarrow Y_2+j$ and $pp\rightarrow Y_2+Z$. Similarly to $pp\rightarrow Y_2$, the very hard tails are seen again in the other two processes, highlighting the unitarity-violating behaviour of the non-universal coupling scenarios. We also separate in Figure~\ref{fig:diff2-b} the contributions from the different helicity configurations of $Y_2$. The leading unitarity-violation behaviour $\hat{s}^3/m_{Y_2}^4/\Lambda^2$  comes from the helicity $h=0$ contribution, the $h=1$ contributions have a subleading growth $\hat{s}^2/m_{Y_2}^2/\Lambda^2$, while $h=2$ curves are consistent with what is expected from dimension-five operators. The dramatic unitarity-violation behaviour of the non-universal coupling case underlines the inadequacy/incompleteness of any naive effective field theory for a massive spin-two particle~\cite{ArkaniHamed:2002sp} and calls for the implementation of extra mechanisms (such as the introduction of other degrees of freedom) that restore unitarity up to scales $\Lambda$ parametrically larger than $m_{Y_2}$.

\section{Partial decay widths}
The LO partial decay widths of the spin-two particle $Y_2$ to SM particles can be written as
\be\bsp
& \Gamma^{\rm LO}(Y_2\rightarrow f\bar{f})=\frac{\kappa_f^2 N_c^f m_{Y_2}^3}{160\pi\Lambda^2}(1-4r_f)^{3/2}(1+\frac{8}{3} r_f),f\neq \nu\\
& \Gamma^{\rm LO}(Y_2\rightarrow \nu_f\bar{\nu}_f)=\frac{\kappa_{\nu_f}^2 m_{Y_2}^3}{320\pi\Lambda^2},\\
\esp\ee

\newpage

\be\bsp
& \Gamma^{\rm LO}(Y_2\rightarrow gg)=\frac{\kappa_g^2 m_{Y_2}^3}{10\pi\Lambda^2},\\
& \Gamma^{\rm LO}(Y_2\rightarrow \gamma\gamma)=\frac{\kappa_{\gamma}^2 m_{Y_2}^3}{80\pi\Lambda^2},\\
& \Gamma^{\rm LO}(Y_2\rightarrow Z\gamma)=\frac{\kappa_{Z\gamma}^2 m_{Y_2}^3}{240\pi\Lambda^2}(1-
r_Z)^3\left(6+3r_Z+r_Z^2\right),\\
& \Gamma^{\rm LO}(Y_2\rightarrow ZZ)=\frac{m_{Y_2}^3}{960\pi\Lambda^2} (1-4r_Z)^{1/2} f(r_Z)\,,\\
& \Gamma^{\rm LO}(Y_2\rightarrow W^+W^-)=\frac{m_{Y_2}^3}{480\pi\Lambda^2} (1-4r_W)^{1/2} f(r_W)\,,\\
& \Gamma^{\rm LO}(Y_2\rightarrow HH)=\frac{\kappa_{H}^2 m_{Y_2}^3}{960\pi\Lambda^2}(1-4r_H)^{5/2},
\esp\ee
where $ f(r_V)=\kappa_H^2+12\kappa_V^2$ 
$+r_V(12\kappa_H^2+80\kappa_H\kappa_V-36\kappa_V^2)$ 
$+r_V^2(56\kappa_H^2-80\kappa_H\kappa_V+72\kappa_V^2)$
,  and we have defined the dimensionless quantities $r_i={m_i^2}/{m_{Y_2}^2}$, $N_c^f$ is the colour of $f$ ({\it i.e.}, $N_c^f=1$ for leptons and $N_c^f=3$ for quarks) and $\kappa_{\gamma}=\kappa_B \cos^2{\theta_W}+\kappa_W\sin^2{\theta_W},\kappa_Z=\kappa_B\sin^2{\theta_W}+\kappa_W\cos^2{\theta_W},\kappa_{Z\gamma}=(\kappa_W-\kappa_B)\cos{\theta_W}\sin{\theta_W}$ with the Weinberg angle $\theta_W$. The above expressions have been checked with \fr\ and have also been numerically validated with \mw~\cite{Alwall:2014bza}. The prefactors $(1-4r_i)^{(2l+1)/2}$ in the massive final state decay modes indicate the (first) angular momentum between the decayed products. For example, due to the fact that the Higgs boson $H$ has a spin of zero, the decay $Y_2\rightarrow HH$ proceeds only through a D-wave, {\it i.e.},  $l=2$. For the partial decay widths of the colored final states (i.e. $Y_2\rightarrow jj$ and $Y_2\rightarrow t\bar{t}$), one can also easily include the NLO corrections within \amc\ framework.

\begin{figure*}[t]
\centering
\vspace{-2cm}
\includegraphics[width=0.7\textwidth]{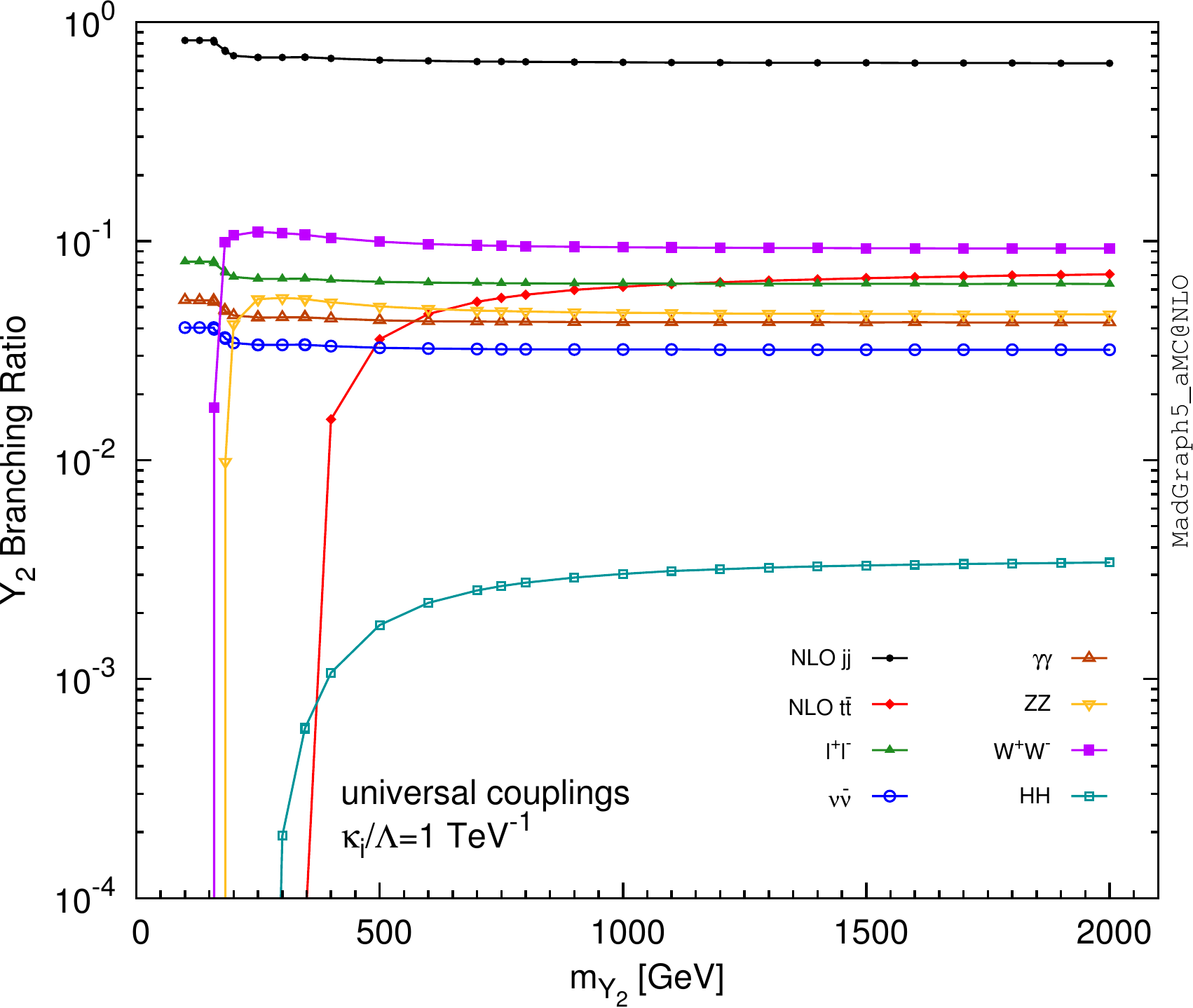}
  \caption{\small \label{fig:width}Branching ratios for the decay modes in the function of the resonance mass $m_{Y_2}$ by assuming the universal coupling case, where NLO QCD corrections are included in the colored final states with the renormalisation scale $\mu_R=m_{Y_2}/2$. Here only the on-shell two-body SM decay modes are considered.}
\end{figure*}

For illustration, we list in Figure~\ref{fig:width} the numerical values for the branching ratios of $Y_2$ in the universal coupling scenario, where the partial widths for $Y_2\rightarrow jj$ and $Y_2\rightarrow t\bar{t}$ include the QCD corrections with the renormalisation scale $\mu_R=m_{Y_2}/2$. The branching ratios are given by assuming that $Y_2$ only decays to SM particles, and by only considering two-body decay modes which are relevant and interesting in experimental new physics searches. In this case, the dominant decay mode is $Y_2\rightarrow jj$. The leptonic decay mode $Y_2\rightarrow \ell^+\ell^-$ (with $\ell^{\pm}=e^{\pm},\mu^{\pm},\tau^{\pm}$ ) is comparable with $Y_2\rightarrow \gamma\gamma$. The partial width of $Y_2\rightarrow HH$ is quite small as it proceeds through a D-wave decay only.

\section{Conclusions}

In summary, we have implemented the Lagrangian of a generic spin-two particle in the \amc\ framework. This allows for the first time to perform simulations of production and decay of a spin-two state $Y_2$ at NLO QCD accuracy  in all the relevant channels at hadron colliders and in particular at the LHC.  They include $Y_2$  production in association with QCD particles (inclusive, one jet, two jets and top quark pair) and with EW bosons ($Z,W^{\pm},\gamma$). In addition, the cross section for the LI process $pp\rightarrow Y_2+H$ is obtained at LO+PS level.  We find that NLO corrections give rise to sizeable $K$ factors for many channels and greatly reduce the theoretical uncertainties in general. All results presented here can be reproduced automatically within the \amc\ framework for both universal coupling and non-universal coupling cases. The unitarity-violation behaviour of the transverse momentum $p_T(Y_2)$ spectra in the non-universal coupling case has been investigated and we find that  the leading (subleading) unitarity-violation term comes from the $h=0$ ($h=1$) amplitudes. Finally, we also presented the general expressions of the LO partial decay widths for $Y_2$ decays into SM particles, and included the NLO QCD corrections for the decays to colored partons, $Y_2\rightarrow jj$ and  $Y_2\rightarrow t\bar{t}$. Our results can be readily used in experimental simulations of spin-two searches and interpretations as well as for other analyses at the LHC, such as the search for spin-two mediators in simplified dark matter models. 

\vspace*{-0.25cm}
\section*{Acknowledgments}
\vspace*{-0.25cm}
\begin{small} 
We are grateful to Stefano Frixione, Christophe Grojean, Prakash Mathews, Kentarou Mawatari for discussions. 
This work has been supported in part by the ERC grant 291377 \textit{LHCtheory: Theoretical predictions and analyses of LHC physics: advancing the precision frontier}, the Research Executive Agency of the European Union
under Grant Agreement PITN-GA-2012-315877 (MCNet). GD is also supported by funding from Department of Atomic Energy, India. CD is a Durham International Junior Research Fellow. The work of VH is supported by the SNSF grant PBELP2 146525.
\\
\end{small}

\bibliographystyle{elsarticle-num}
\bibliography{huasheng}

\end{document}